\documentclass[twocolumn,showpacs,preprintnumbers,groupedaddress,amsmath,amssymb,floatfix,prl]{revtex4}
\usepackage{graphicx}
\usepackage{epsfig}

\begin{document}


\title{Using Three-Body Recombination to Extract Electron Temperatures of Ultracold Plasmas}


\author{R. S. Fletcher, X. L. Zhang, and S. L. Rolston}
\affiliation{Joint Quantum Institute, Department of Physics, University of Maryland, College Park, MD 20742}


\date{\today}

\begin{abstract}
Three-body recombination, an important collisional process in plasmas, increases dramatically at low electron temperatures, with an accepted scaling of T$_{\rm e}^{-9/2}$.  We measure three-body recombination in an ultracold neutral xenon plasma by detecting recombination-created Rydberg atoms using a microwave-ionization technique.  With the accepted theory (expected to be applicable for weakly-coupled plasmas) and our measured rates we extract the plasma temperatures, which are in reasonable agreement with previous measurements early in the plasma lifetime.  The resulting electron temperatures indicate that the plasma continues to cool to temperatures below 1 K.
\end{abstract}


\maketitle

Three-body recombination ($e^- +e^- + A^+ \rightarrow e^- +A^* $) is a fundamental collisional process in plasmas that is  dominant at sufficiently low electron temperatures due to its $ T{\rm _e^{-9/2}}$ dependence. In ultracold plasmas (UCPs), the observation of copious  Rydberg atom production \cite{killian2001} and the observation of $T_{\rm e}$ almost independent of initial energies \cite{roberts2004} show that three-body recombination (3BR) and its associated heating play a critical role in the evolution of UCPs.  Early-time $T_{\rm e}$ measurements\cite{roberts2004} and simulations\cite{commRobicheaux} suggest that the electrons remain weakly coupled, so that traditional 3BR theory is still valid (in the strong-coupling regime, the 3BR rate is predicted to be reduced below the 9/2 scaling law to a $T{\rm _e^{-1}}$ rule \cite{hahn}). A measurement of 3BR in an UCP can thus be used to test 3BR theory by using existing $T_{\rm e}$ measurements. This is less than ideal, given the paucity of measurements and the sensitivity of the rate constant to $T_{\rm e}$ due to the 9/2 power.  Conversely, using 3BR theory, $T_{\rm e}$ can be extracted from the measured 3BR rate.  This is relatively insensitive to the value of the rate constant (due to a 2/9 power law), and can be used to make the first measurement of  $T_{\rm e}$  throughout the life of the plasma. In addition, modifications of the rate constant due to strong-coupling will overestimate $T_{\rm e}$ , i.e. our extracted $T_{\rm e}$ are an upper limit.   We note that in addition to  furthering our  understanding of UCPs, a study of 3BR may aid in using plasmas with similar parameters (albeit at high magnetic fields)  to optimize production of anti-hydrogen \cite{gabrielse}.


In this work, we measure the instantaneous Rydberg atom production rate in an expanding  ultracold xenon plasma  as a function of the time elapsed after plasma formation.  By applying a short microwave pulse to the UCP, the Rydberg population for principal quantum numbers N$>$35 is ionized and subsequently detected by a micro-channel plate detector.  Using two such pulses separated by up to a few microseconds, we measure the refilling of the depleted Rydberg levels that form during the interval between the two pulses. In this manner we determine the instantaneous  Rydberg formation rate  during the plasma expansion.  There are several processes that may contribute to this refill rate, including 3BR, radiative recombination, blackbody-driven transitions from low Rydberg levels to higher (and thus microwave-ionizable) levels, and electron-Rydberg collisions that drive transitions between N-levels.  Of these processes, 3BR has the strongest inverse-temperature dependence and will be the most prevalent process at low temperatures.  Simulations based on the accepted expressions for these various processes and our $T_{\rm e}$ support our assumption that 3BR is the only significant process, so that the refill rates we observe are due to 3BR.  We measure the total 3BR rate, which should be proportional to $ K_{\rm{3BR}} T{\rm _e^{-9/2}}$, where $ K_{\rm{3BR}}$ is the three-body rate constant.  We use the accepted $ K_{\rm{3BR}}$ to extract $T_{\rm e}$ at later times in the plasma, yielding the first $T_{\rm e}$ measurements in an UCP later than 10 $\mu$s after the plasma creation. 
  
Our creation of the UCP is similar to previous work \cite{killian1999}.  A magneto-optic trap  is used to collect $\sim 5 \times 10^6$ metastable xenon atoms and to cool them to a temperature of approximately 20 $\mu$K.  The density distribution is roughly Gaussian with a rms radius $\sigma_0\sim 260\,\mu$m and a peak density of about $ n_0  = 1 \times 10^{16} \,{\rm m^{-3}}$.  The plasma is then produced using a two-photon excitation process (882 nm + 514 nm, 10ns pulse), ionizing $\sim$  30$\%$ of the atoms. The initial electron energy $\Delta$E is controlled by the frequency of the 514-nm photon; for this work we use $\Delta$E $=$ 3 K.  Using a micro-channel plate detector and a series of biased voltage grids (the plasma is in a $\sim$ 5 mV/cm electric field), we detect the emission of electrons from the plasma. Figure 1 is an example of such a signal.

The plasma is unconfined, expands freely into vacuum, and lives for $\sim$ 200 $\mu$s.  As it expands, it maintains a roughly Gaussian density distribution (confirmed by ion imaging, to be published); we assume that
\begin{equation}
n_e(r,t) \sim n_i(r,t) \sim \frac{n_0\sigma_0^3}{\sigma(t)^3}exp[-r^2/2\sigma(t)^2]
\end{equation}
where $ \sigma(t)=\sqrt{\sigma_0^2 + v^2t^2}, \sigma_0 \sim 260 \, \mu {\rm m}, v \sim 65\, {\rm m/s}$, and $ n_0 \sim 2\times10^{15}\, {\rm m^{-3}}$ \cite{killian2004, bergeson2003, kulin2000}.  While simulations indicate some deviation from this distribution \cite{robPlas, pohlModel}, we note that this deviation should have only a small effect on our application of eq. 1.  Furthermore, although the plasma is constantly losing both ions and electrons, $n_{\rm e}$ and $ n_{\rm i}$  will remain approximately equal, especially at the plasma center.  Most deviations from the Gaussian distribution occur in the outer, low-density region, but since almost all 3BR occurs in the high-density central region of the plasma due to the $n_e^2 n_i$ dependence of the 3BR rate:  
\begin{equation}
R_{3BR}\,(s^{-1}) = K_{3BR}Ê\, T_e^{-9/2}\int{n_e^2(r)n_i(r)4\pi r^2dr}, 
\end{equation}
eq. 1 is a good approximation for use in 3BR calculations.  

In addition to this $t$- and $r$-dependent density distribution, the plasma also exhibits time-dependent electron and ion temperatures due to heating effects (3BR, disorder-induced heating) and cooling effects (adiabatic expansion, electron evaporation).  Effective ion temperatures (distinct from thermal temperatures) have been directly measured \cite{killian2004, bergeson2005}, but only during the first $\sim$ 4 $\mu$s of plasma expansion.  While the thermal electron temperature, $T_{\rm e}$, has been estimated using simulation results \cite{robPlas, pohlModel}, measurements have only been made for the first 10 $\mu$s of plasma expansion \cite{roberts2004}.  The use of Tonks-Dattner resonances as a plasma diagnostic may provide a way to measure $T_{\rm e}$ at times up to $\sim$ 35 $\mu$s \cite{fletcher}; however,  current  theory prevents it from being a quantitative measurement.  Our 3BR measurements span a much larger region of the plasma lifetime than $T_{\rm e}$ measurements, allowing only limited comparison of our measured rates to calculated 3BR rates using known $T_{\rm e}$.  We  note that significant Rydberg production occurs throughout the entire lifetime of the plasma \cite{killian2001}, allowing temperature information to be determined throughout.

\begin{figure}[htbp]
\vskip -.0in
\begin{center}
\epsfig{file=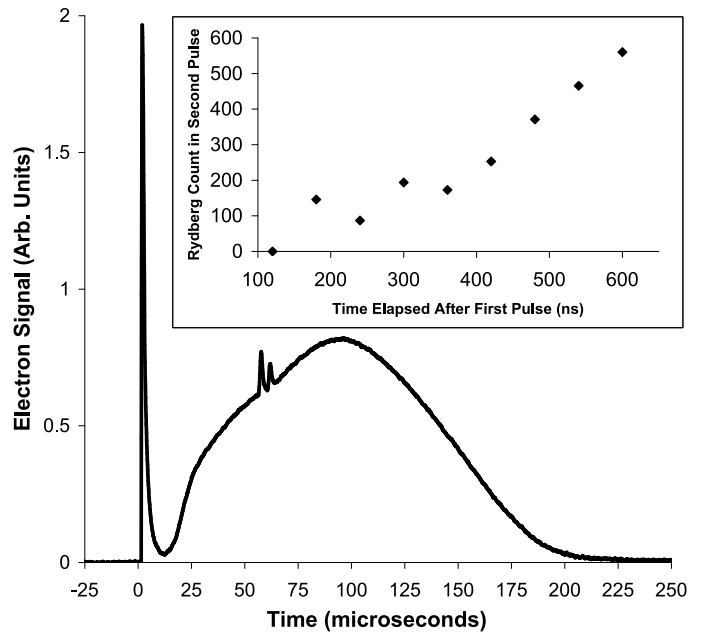, width=3.53in}
\end{center}

\vskip -.2in
\caption{ Electron emission signal from an expanding UCP, averaged over 40 runs.  The double peaks at $\sim 55  \mu$s are the response due to a pair of short (100 ns) microwave pulses.  The first of the two pulses is held fixed and ionizes Rydbergs that have formed in the plasma.  Following this pulse, the Rydberg population is refilled by three-body recombination.  The second pulse is applied at varying times and the number of Rydbergs ionized is counted as a function of the time between the two pulses (inset).  }
\end{figure}

Microwave ionization of Rydbergs is a non-resonant process and largely independent of the microwave frequency; we use 2.4 GHz pulses to take advantage of an electrical resonance in our apparatus.  We find that these pulses ionize Rydbergs with N$>$35 by applying a DC field ionization ramp immediately after a microwave pulse and noting that the signal corresponding to N$>$35 Rydbergs is almost entirely eliminated.  We estimate a microwave field amplitude of $\sim$ 220 V/cm by observing which principle quantum numbers are ionized, as we are unable to directly measure the microwave field at the plasma. We find that we are not directly heating the plasma electrons.  Starting with a plasma electron energy of  $\Delta$E= 300 K (to suppress Rydberg formation) and applying the microwave pulses at varying times in the plasma, we observe that the plasma expansion and the electron emission signal do not substantially change with the addition of the microwave pulses.  The microwave pulses can heat the plasma indirectly through the ionization electrons, as those electrons undergo a few collisions while exiting the plasma. The electrons resulting from microwave ionization are warm ($>$10 K) \cite{commRobicheaux} with respect to the plasma electrons, but leave the plasma quickly, with limited thermalization with the plasma electrons.  We minimize this effect by keeping the microwave pulse duration as short as possible (100 ns) while still fully ionizing the Rydberg population.  

We apply the first microwave pulse at a chosen time in the plasma expansion in order to remove the existing N$>$35 population.  We then apply a second pulse after a variable delay.  This results in a double-peaked electron signal (fig. 1 near 55 $\mu$s).  We fit to the broad emission curve and use this as a background to determine the number of electrons contained in the second peak\cite{numcalibration}.  A single microwave pulse (as well as DC field ionization) can also provide a means to measure the number of Rydberg atoms in the plasma, but the long Rydberg lifetimes and the collisional redistribution of Rydberg states make this measurement difficult to interpret in the context of 3BR, as it is the result of past Rydberg production and redistribution.  

We note that the width of a electron emission peak due to our short microwave pulse is a few times wider than the duration of the microwave pulse.  There is also a delay before we begin to detect Rydberg refilling (fig. 1 inset).  If the plasma is heated a small amount, as is likely following a microwave ionizing pulse, 3BR would be inhibited immediately after an ionizing pulse, explaining the delay before we observe Rydberg formation.  This is not a major effect, however, because the energy put into the plasma electrons by the newly-ionized electrons should largely be removed through the loss of hot electrons (as observed in our electron emission peaks).  The increased width of the electron emission peaks is likely the result of a Coulomb-driven spread of the electron bunch during the time of flight to the detector.

Using this two-pulse method,  we obtain the Rydberg atom refill rate throughout the plasma expansion.  A typical refill curve (for t = 55 $\mu$s) is shown in the inset plot of figure 1.  We fit a straight line to the first few hundred nanoseconds of each refill curve and use the slope of that line as the short-time refill rate (that is, the instantaneous 3BR rate).  At late times ($>$ 1-2 $\mu$s) in the refill curve the rate decreases, likely due to the plasma regaining a near-equilibrium Rydberg distribution.  

Figure 2 shows the measured Rydberg refill rates.  Because multiple processes can create Rydbergs in the N$>$35 states, we considered the possibility that these measured rates may be dominated by non-3BR events.  Using  established rates for radiative recombination \cite{seaton}, blackbody-driven transitions and ionization \cite{gallagher}, radiative decay \cite{gallagher}, and 3BR \cite{mansbach}, we simulate\cite{monte} the change in the populations in each $N$ level for 3$<$N$<$200 for several $\mu$s starting at each time we obtained refill rates, using the plasma densities at that time (eq. 1) with varied $T_{\rm e}$.  We use an initial Rydberg population distribution which is truncated at N$=$35 to simulate the microwave ionization event, then include all of the above-cited processes in a Monte-Carlo  model of the time-dependent Rydberg distribution.   The results of this simulation confirm that 3BR is the process most responsible for creating Rydbergs in N$>$35 states following an ionization pulse, as the other processes together contribute less than a few percent to the rates we measure.  The simulation also supports our measurement of only N$>$35 Rydbergs as a complete 3BR measurement, as 3BR at such low temperatures almost exclusively populates high  levels (N$\geq$ 80).  

\begin{figure}[htbp]
\vskip -.0in
\begin{center}
\epsfig{file=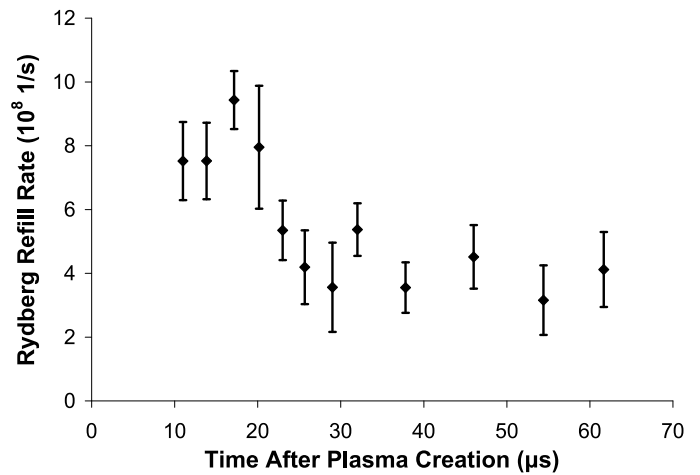, width=3.53in}
\end{center}

\vskip -.2in
\caption{ The total Rydberg refill rates.  The early increase in the rate is likely due to 3BR turning on as temperatures drop; the rate peaks and then drops to a constant value as the density and temperature continue to decrease. Note that the peak plasma density falls by over two orders of magnitude, from $1 \times$10$^{14}$ m$^{-3}$ at 10 $\mu$s to $6 \times$10$^{11}$ m$^{-3}$ at 60 $\mu$s, while the 3BR rate only decreases by a factor of two.  The error bars represent  the 1$\sigma$ standard uncertainty resulting from the linear fits to the refill curves (such as the inset of fig. 1).}
\end{figure}

If we then assume that 3BR theory with $K_{\rm 3BR} =4.5 \times 10^{-21} {\rm m^{6} K^{9/2} s^{-1}}$ provides us with an accurate rate equation (eq. 2) \cite{mansbach}, we can use our measured recombination rates together with the Gaussian expansion approximation (eq. 1) to extract  $ T_{\rm e}(t)$, plotted in fig. 3a.  We note that these $ T_{\rm e}(t)$ are similar to those predicted by simulations \cite{robPlas, commRobicheaux}.  A simple power-law fit of the temperature curve of figure 3a gives $ T_{\rm e}(t) \sim t^{-1.2(1)}$.  Adiabatic cooling would suggest a  $ T_{\rm e}(t) \sim t^{-2}$ relationship; the difference  is likely due to the significant heating effects from 3BR.  This method of determining the electron temperature finds that the electrons are well below one Kelvin at later times, by far the lowest $T_{\rm e}$ observed in an ultracold plasma.  Indeed, extrapolating to the end of the plasma lifetime ($\sim$ 200 $\mu$s), the power-law fit would indicate temperatures as low as $\sim$ 200 mK.  While we observe Rydberg production  throughout the lifetime of the plasma, we did not measure  past t=65 $\mu$s because the indirect heating effects of the 100 ns microwave pulses become significant compared to $T_{\rm e}$, interfering with extraction of the 3BR rates.    We note that $T_{\rm e}$ is rather insensitive to the value of $K_{\rm 3BR}$, as a factor of two change in the rate constant results in only a $16 \% $ change in temperature.

The temperature in figure 3a is for a plasma with an initial $T_{\rm e}$ of 3 K.  Also plotted are previous $T_{\rm e}$ measurements for an initial $\Delta$E=10 K (squares, \cite{roberts2004}) and simulation results for a $\Delta$E=66 K (triangles, \cite{commRobicheaux}).  Despite the different initial temperatures, the three sets are roughly consistent with one another; as seen in \cite{roberts2004}, the $T_{\rm e}$ tend to converge to similar values despite very different $\Delta$E's.  

Using the temperature calculations of figure 3a with the density expression given in eq. 1, we calculate the electron coupling parameter $\lambda=e^2 /(4\pi\epsilon_0 ak_b T_e$), where $a$ is the average Wigner-Seitz length.   $\lambda (t)$  is plotted in figure 3b.  The system remains weakly coupled ($\lambda \ll 1$), but the coupling is increasing as the plasma expands and cools.  

Some work has been done to modify eq. 2.  Using a Michie-King distribution for the electron velocity distributions instead of a Maxwell-Boltzmann distribution results in an average corrective factor to eq. 2 of 1 to 1.1 \cite{pohlTBR},  increasing our temperatures less than 2.5 $\%$.  Modifications to eq. 2 to correct  for low temperatures by using a density dependent cutoff for the Rydberg levels created by 3BR \cite{hahn} do not apply to our plasma.  The theory modifications are only applicable for a plasma denser than ours by several orders of magnitude (our plasma is still in the weakly-coupled regime). Any strong-coupling correction is likely to result in lower calculated 3BR rates at a particular $T_{\rm e}$.  This indicates that if such a correction to 3BR rates is needed for the parameters of this UCP, our calculated $T_{\rm e}$ would be an {\it upper} bound estimate of the actual plasma $T_{\rm e}$.  It should be noted that even a substantial change to the 3BR rate equation could be masked in the temperature plot of figure 3a due to the exponent of 9/2 on the $T_{\rm e}$ term.  

Recent work \cite{Hu} indicates that at electron temperatures below $\sim$ 1000 K, quantum 3BR should populate lower N levels than classical 3BR theory would predict.  Classical 3BR theory predicts almost no formation in N$<$35 states. If quantum 3BR effects result in the formation of N$<$35 Rydbergs, the rates we measure in this work would undercount the actual 3BR rate.  We can not directly compare our results with the calculations presented in \cite{Hu}, as those calculations did not go lower than $T_{\rm e}$ $\sim$ 100 K, nor have we made comprehensive measurements of the N-level distributions that are created in the 3BR-driven refilling process.  However, our apparent agreement with classical 3BR calculations suggests that quantum 3BR at $T_{\rm e}$ $\sim$ 1-10 K has little effect on the cumulative N$>$35 population.

\begin{figure}[htbp]
\vskip -.0in
\begin{center}
\epsfig{file=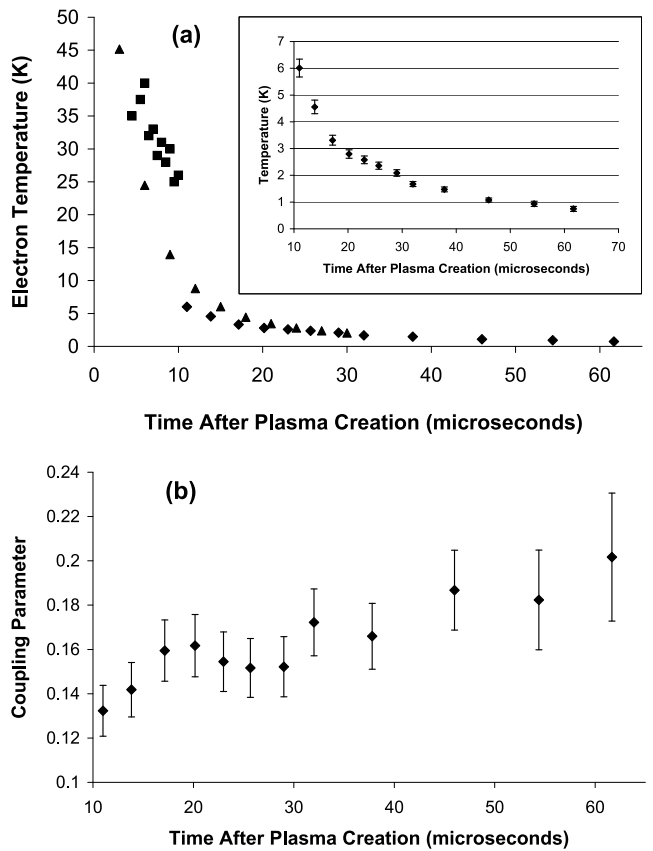, width=3.53in}
\end{center}

\vskip -.2in
\caption{ (a)  We calculate $T_{\rm e}$ using established 3BR theory with the rates we measure and an assumed self-similar Gaussian expansion of the plasma (diamonds; also inset).  The initial energy  is $\Delta$E=3 K.  For comparison, we plot earlier $T_{\rm e}$ measurements for $\Delta$E=10 K (squares) \cite{roberts2004} and simulation results for $\Delta$E=66 K (triangles) \cite{commRobicheaux}. (b)  The corresponding plasma coupling parameter $\lambda$.    The error bars are the propagated  uncertainties  from figure 2.}
\end{figure}

We have directly observed the Rydberg formation rate at varying times in the expansion of an ultracold plasma by using a short microwave pulse to ionize a Rydberg population, followed by a second pulse to probe the refilling of that population.  Simulations show that 3BR is the mechanism by which this Rydberg population refills, because other mechanisms are suppressed due to the low temperatures in an ultracold plasma.  In future work, we can measure the  N-level distribution that is formed by 3BR to determine the importance of quantum effects on 3BR theory \cite{Hu}, by varying the microwave field strength of the second pulse.

Although standard 3BR theory is expected to fail in strongly coupled systems, we use a typical 3BR expression and extract $T_{\rm e}(t)$ for the plasma.  Because the resulting $T_{\rm e}$ estimates match well with another measurement of the temperature and with simulation results (at least at early times in the plasma expansion), the use of the accepted 3BR rate expression in this work appears justified.  Our method is self-consistent in that the extracted temperatures show the plasma to still be weakly coupled. Future  measures of $T_{\rm e}$ in an independent manner  would allow a direct test the validity of the 3BR expression in UCPs.  This is the lowest-$T_{\rm e}$ measurement of three-body recombination rates  to date; the measured rates indicate the plasma system achieves sub-Kelvin electron temperatures at late times.

\begin{acknowledgments}
This work was partially supported by the National Science Foundation PHY0245023.
\end{acknowledgments}

\bibliography{TD}

\begin{thebibliography}{20}
\bibitem[{num()}]{numcalibration}
\bibinfo{note}{The calibration of the electron signal due to the second pulse
  is determined as a ratio to the total plasma signal, which is calibrated from
  atom loss measured with laser absorption. This yields a systematic 10$\%$
  uncertainty to the measurements.}
\bibitem[{mon()}]{monte}
\bibinfo{note}{The Monte Carlo simulation is similar to that of ref. [10].}
\end{thebibliography}

\end{document}